\documentstyle[takayama,12pt]{article}   
\input{epsf}

\renewcommand{\theequation}{\arabic{section}.\arabic{equation}}

\begin{document}

\refNP
\final
                                      
\date{TIT/HEP-406/NP\\
      November, 1998}
                                      
\title{Hyperon Non-leptonic Weak Decays in the Chiral Perturbation
Theory II}

\author{K.~Takayama and M.~Oka
        \affiliation{Department of Physics, 
                    Tokyo Institute of Technology\\
             Meguro, Tokyo 152, Japan}  }

\maketitle

\abstract{
Hyperon non-leptonic weak decay amplitudes are studied in the chiral
perturbation theory. The weak interaction vertices caused by the four
quark operators are substituted by the products of the hadronic currents 
and by the phenomenologically introduced weak Hamiltonian of hadron
operators. Our study suggests the improvement of the theoretical
prediction for the weak decay amplitudes.
}

\thispagestyle{empty}
\newpage
\setcounter{equation}{0}
\setcounter{figure}{0}
\section{Introduction}
The chiral perturbation theory is applied to many hadron
phenomena, and succeed to reproduce the experimental data. This theory
is applied to the analysis of hyperon non-leptonic weak decays.
But this theory cannot reproduce the
experimental data\cite{Bijnens85,Jenkins92b}. In the previous
paper\cite{Takayama98}, we use the chiral
perturbation theory as the non-perturbative QCD correction for the weak 
interaction. The quark four point vertices are substituted by the
product of the hadron currents which is derived by the chiral
perturbation theory. We can reproduce the suppression of the
$\Delta I=3/2$ amplitudes. But this method cannot reproduce the
$\Delta I=1/2$ amplitudes. It suggests that the product of the hadron
currents cannot reproduce all the four point vertices of the quark
operators. In this study, we construct the effective weak Hamiltonian
which cannot be represented by the product of the hadron currents, and 
consider the role of the effective Hamiltonian.

This paper is organized as follows. In section 2 we construct the
effective weak Hamiltonian. Section 3 presents the numerical analysis
and discussion on the hyperon non-leptonic weak decay
amplitudes. Section 4 concludes the paper with the comment on the
effective weak Hamiltonian using the chiral perturbation theory.
\setcounter{equation}{0}
\setcounter{figure}{0}
\section{The effective weak Hamiltonian}
\subsection{The current-current interaction}

In order to apply the standard theory to the hyperon non-leptonic weak 
decay, the momentum transfer scale of the weak interaction vertex is
changed with the renormalization group method. This method constructs
the effective weak Hamiltonian including the perturbative QCD
correction to the weak interaction\cite{vzs77,Paschos90}. This Hamiltonian is
represented by the four point vertices of the quark operator.

Applying the four quark vertex to the hyperon non-leptonic weak decay, 
we can derive two types of the diagrams, which are shown in
Fig.~2.1. In order to derive the effective weak Hamiltonian with the
chiral perturbation theory, the quark lines, which have color index,
are substituted by the hadron operators which are color singlet. The
diagram in Fig.~2.1(a) can be substituted by the product of the hadron
currents, but the diagram in Fig.~2.1(b) cannot be represented by
it. For the construction of the effective weak Hamiltonian of the
diagram in Fig.~2.1(b), we introduce the two ansatz. 
\begin{itemize}
  \begin{enumerate}
    \item Applying Fierz transformation to the diagrams in Fig.~2.1,
          the diagram (b) is changed to the product of the color singlet
          quark currents. And this product is substituted by the
          product of the hadron currents.
    \item For the residual effect of the ansatz~1, we introduce the
          phenomenological effective weak Hamiltonian which has
	  parameters. 
  \end{enumerate}
\end{itemize}
In the following, the effective weak Hamiltonian described by the
product of the hadron currents is called as the current-current
interaction, and the effective weak Hamiltonian derived by the
ansatz~2 is called as the internal interaction. 

Using the above ansatz and the heavy baryon formalism of the chiral
perturbation theory, the effective weak Hamiltonian is given by
\begin{equation}
 {\cal H}^{\Delta S=1}_{eff}
      ={\cal H}_{current}^{\Delta S=1}
      +{\cal H}_{int}^{\Delta S=1}
   \label{eqn:hamil1}
\end{equation}
\begin{equation}
 {\cal H}^{\Delta S=1}_{current} =
     -\frac{G_f}{\sqrt{2}}\sum_{r=1}^{6}K_rO_r,
  \label{eqn:weak1}
\end{equation}
where the coefficients $K_r$ is shown in Table~2.1. The operators
$O_r$ are given by
\begin{equation}
  \begin{array}{lcl}
  O_1 &=& \frac{2}{3}\left({\cal J}^{23}_{L\mu}{\cal J}^{11\mu}_{L}
         -{\cal J}^{13}_{L\mu}{\cal J}^{21\mu}_{L}\right)
  \\
  O_2 &=& \frac{4}{3}\left({\cal J}^{23}_{L\mu}{\cal J}^{11\mu}_{L}
         +{\cal J}^{13}_{L\mu}{\cal J}^{21\mu}_{L}
         +2{\cal J}^{23}_{L\mu}{\cal J}^{22\mu}_{L}
         +2{\cal J}^{23}_{L\mu}{\cal J}^{33\mu}_{L}\right)
  \\
  O_3 &=& O_3\left(\Delta I=1/2 \right)
  \\
      &=& \frac{4}{9}\left({\cal J}^{23}_{L\mu}{\cal J}^{11\mu}_{L}
         +{\cal J}^{13}_{L\mu}{\cal J}^{21\mu}_{L}
         +2{\cal J}^{23}_{L\mu}{\cal J}^{22\mu}_{L}
         -3{\cal J}^{23}_{L\mu}{\cal J}^{33\mu}_{L}\right)
  \\
  O_4 &=& O_3\left(\Delta I=3/2 \right)
  \\
      &=& \frac{20}{9}\left({\cal J}^{23}_{L\mu}{\cal J}^{11\mu}_{L}
         +{\cal J}^{13}_{L\mu}{\cal J}^{21\mu}_{L}
         -{\cal J}^{21}_{L\mu}{\cal J}^{22\mu}_{L}\right)
  \\
  O_5 &=& O_{51}+O_{52}+O_{53}
  \\
  &&O_{51} = \left({\cal J}^{23}_{L\mu}{\cal J}^{11\mu}_{R}
         +{\cal J}^{23}_{L\mu}{\cal J}^{22\mu}_{R}
         +{\cal J}^{23}_{L\mu}{\cal J}^{33\mu}_{R}\right)
  \\
  &&O_{52} = -\frac{2}{3}\left({\cal J}^{21}_{0(S+iP)}{\cal J}^{13}_{0(S-iP)}
         +{\cal J}^{22}_{0(S+iP)}{\cal J}^{23}_{0(S-iP)}
         +{\cal J}^{23}_{0(S+iP)}{\cal J}^{33}_{0(S-iP)}\right)
  \\
  &&O_{53} = -\frac{2}{3}\left({\cal J}^{21}_{2(S+iP)}{\cal J}^{13}_{0(S-iP)}
         +{\cal J}^{21}_{0(S+iP)}{\cal J}^{13}_{2(S-iP)}
         +{\cal J}^{22}_{2(S+iP)}{\cal J}^{23}_{0(S-iP)}\right.
  \\
         &&\left. \qquad\quad
         +{\cal J}^{22}_{0(S+iP)}{\cal J}^{23}_{2(S-iP)}
         +{\cal J}^{23}_{2(S+iP)}{\cal J}^{33}_{0(S-iP)}
         +{\cal J}^{23}_{0(S+iP)}{\cal J}^{33}_{2(S-iP)}\right)
  \nonumber\\
  O_6 &=& O_{61}+O_{62}
  \\
  &&O_{61} = -\frac{32}{9}\left({\cal J}^{21}_{0(S+iP)}{\cal J}^{13}_{0(S-iP)}
         +{\cal J}^{22}_{0(S+iP)}{\cal J}^{23}_{0(S-iP)}
         +{\cal J}^{23}_{0(S+iP)}{\cal J}^{33}_{0(S-iP)}\right)
  \\
  &&O_{62} = -\frac{32}{9}\left({\cal J}^{21}_{2(S+iP)}{\cal J}^{13}_{0(S-iP)}
         +{\cal J}^{21}_{0(S+iP)}{\cal J}^{13}_{2(S-iP)}
         +{\cal J}^{22}_{2(S+iP)}{\cal J}^{23}_{0(S-iP)}\right.
  \\
         && \left. \qquad\quad
         +{\cal J}^{22}_{0(S+iP)}{\cal J}^{23}_{2(S-iP)}
         +{\cal J}^{23}_{2(S+iP)}{\cal J}^{33}_{0(S-iP)}
         +{\cal J}^{23}_{0(S+iP)}{\cal J}^{33}_{2(S-iP)}\right)
  \end{array}
  \label{eqn:operator1}
\end{equation}
\begin{equation}
  \begin{array}{lcl}
 {\cal J}^{ij}_{L\mu}
  &=&\mbox{Tr}v_{\mu}\bar{B}_v\left[\xi^{\dag}h_{ij}\xi,B_v \right]
  -2D\mbox{Tr}\bar{B}_v S_{v \mu}\left\{\xi^{\dag}h_{ij}\xi,B_v \right\}
   \\
  &&-2F\mbox{Tr}\bar{B}_v S_{v \mu}\left[\xi^{\dag}h_{ij}\xi,B_v\right]
  \\
  &&-v_{\mu}\bar{T}^{\nu}_v\left( \xi^{\dag} h_{ij}\xi \right)T_{v\nu}
  -2{\cal H}\bar{T}^{\nu}_v S_{v\mu} \left( \xi^{\dag}h_{ij}\xi \right)T_{v\nu}
   \\
  &&-2{\cal C}\left(\bar{T}_{v\mu} \left(\xi^{\dag} h_{ij}\xi\right) B_v
     +\bar{B}_v \left(\xi^{\dag} h_{ij}\xi \right) T_{v\mu}\right)
  \\
  &&+if^2_{\pi}\mbox{Tr}
        \left(h_{ij}(\partial_{\mu}\Sigma)\Sigma^{\dag}\right),
  \end{array}
  \label{eqn:left-c}
\end{equation}
\begin{equation}
  \begin{array}{lcl}
 {\cal J}^{ij}_{R\mu}
  &=&\mbox{Tr}v_{\mu}\bar{B}_v\left[\xi h_{ij}\xi^{\dag},B_v \right]
  +2D\mbox{Tr}\bar{B}_v S_{v \mu}\left\{\xi h_{ij}\xi^{\dag},B_v \right\}
   \\
  &&+2F\mbox{Tr}\bar{B}_v S_{v \mu}\left[\xi h_{ij}\xi^{\dag},B_v\right]
  \\
  &&-v_{\mu}\bar{T}^{\nu}_v\left( \xi h_{ij}\xi^{\dag} \right)T_{v\nu}
  +2{\cal H}\bar{T}^{\nu}_v S_{v\mu} 
   \left( \xi h_{ij}\xi^{\dag} \right)T_{v\nu}
   \\
  &&+2{\cal C}\left(\bar{T}_{v\mu} \left(\xi h_{ij}\xi^{\dag}\right) B_v
     +\bar{B}_v \left(\xi h_{ij}\xi^{\dag} \right) T_{v\mu}\right)
  \\
 &&+if^2_{\pi}\mbox{Tr}\left(h_{ij}(\partial_{\mu}\Sigma^{\dag})
    \Sigma\right),
  \end{array}
  \label{eqn:right-c}
\end{equation}
\begin{eqnarray}
 {\cal J}_{(S+iP)}^{ij} =
       {\cal J}_{0(S+iP)}^{ij} &+& {\cal J}_{2(S+iP)}^{ij} =
  \\
  {\cal J}_{0(S+iP)}^{ij} &=&
  a_1\mbox{Tr}\bar{B}_v\left\{\xi^{\dag}h_{ij}\xi^{\dag},B_v\right\}
  +a_2\mbox{Tr}\bar{B}_v\left[\xi^{\dag}h_{ij}\xi^{\dag},B_v\right]
  \nonumber\\
  &&+a_3\mbox{Tr}\left(\bar{B}_vB_v\right)
   \mbox{Tr}\left(\xi^{\dag}h_{ij}\xi^{\dag}\right)
  \nonumber\\
  &&+c_1\bar{T}^{\mu}_v \left(\xi^{\dag}h_{ij}\xi^{\dag}\right)T_{v\mu}
    +c_2\bar{T}^{\mu}_vT_{v\mu}
   \mbox{Tr}\left(\xi^{\dag}h_{ij}\xi^{\dag}\right)
   \\
  &&+\frac{f^2_{\pi}}{2}B_0\mbox{Tr}\left(h_{ij}\Sigma^{\dag}\right)
   \nonumber\\
  {\cal J}_{2(S+iP)}^{ij} &=& 2B_0L_4
   \mbox{Tr}\left({\cal D}_{\mu}\Sigma{\cal D}^{\mu}\Sigma^{\dag}\right)
    \mbox{Tr}\left(h_{ij}\Sigma^{\dag}\right)
  \nonumber\\
  &&+2B_0L_5\mbox{Tr}\left(h_{ij}
      {\cal D}_{\mu}\Sigma{\cal D}^{\mu}\Sigma^{\dag}\Sigma^{\dag}\right),
  \label{eqn:plus-c}
\end{eqnarray}
\begin{eqnarray}
 {\cal J}_{(S-iP)}^{ij} =
   {\cal J}_{0(S-iP)}^{ij} &+& {\cal J}_{2(S-iP)}^{ij}
  \\
  {\cal J}_{0(S-iP)}^{ij} &=& 
   a_1 \mbox{Tr}\bar{B}_v \bigg\{ \xi h_{ij}\xi,B_v \bigg\}
  +a_2 \mbox{Tr}\bar{B}_v \bigg[\xi h_{ij}\xi,B_v \bigg]
  \nonumber\\
  &&+a_3 \mbox{Tr} \bigg( \bar{B}_vB_v \bigg)
   \mbox{Tr} \bigg( \xi h_{ij}\xi \bigg)
  \nonumber\\
  &&+c_1\bar{T}^{\mu}_v \bigg(\xi h_{ij}\xi\bigg)T_{v\mu}
    +c_2\bar{T}^{\mu}_vT_{v\mu}
   \mbox{Tr}\bigg(\xi h_{ij}\xi\bigg)
  \\
  &&+\frac{f^2_{\pi}}{2}B_0\mbox{Tr}\left(h_{ij}\Sigma\right)
  \nonumber\\
 {\cal J}_{2(S-iP)}^{ij} &=&
   2B_0L_4
   \mbox{Tr}\left({\cal D}_{\mu}\Sigma{\cal D}^{\mu}
           \Sigma^{\dag}\right)\mbox{Tr}\left(h_{ij}\Sigma\right)
  \nonumber\\
  &&+2B_0L_5\mbox{Tr}\left(h_{ij}
   {\cal D}_{\mu}\Sigma{\cal D}^{\mu}\Sigma^{\dag}\Sigma\right).
  \label{eqn:minus-c}
\end{eqnarray}
The flavor changing operator $h_{ij}$ is given by
\begin{equation}
  \left(h_{ij}\right)_{ab}=\left\{
  \begin{array}{ll}
  1&a=i\;\mbox{and}\;b=j\\
  0&\mbox{others.}
  \end{array}
  \right.
  \label{eqn:matrix-h}
\end{equation}
In the following, we consider the effective weak Hamiltonian for the
internal interaction.

\subsection{Internal interaction}

The diagram in Fig.~2.1(b) suggests that the interaction is described
by the two point baryon vertex. And it has the operator changing the
flavor $s \longrightarrow d$ for the $|\Delta S|=1$ weak
decay. Therefor the effective weak Hamiltonian of the internal
interaction is the extension of the two point baryon vertex. And this
Hamiltonian has the flavor changing operator $h_{23}$ between the
baryon operators. If the internal interaction Hamiltonian depends on the
meson momentum, the lowest order of the weak vertex becomes the three
point vertex constructed by the one meson and two baryon operators. We 
adopt only the momentum independent form for the internal interaction
Hamiltonian in this study. The momentum dependent weak Hamiltonian is
included in the current-current interaction.

The quark level effective weak Hamiltonian has six
type operators, $O_1,\cdots, O_6$. If the internal interaction is
considered outside the baryon, we can observe only the net flavor
changing process $\left(\bar{d}s\right)$ and we cannot distinguish
six operators. The internal interaction Hamiltonian, therefore, is
described by the summation of the six operators.
It is known that the weak interaction has only
$V-A$ type symmetry\cite{weak}. This symmetry is represented by the
transformation symmetry of the operator $h_{23}$.
But in our method,
Fierz transformation is applied to the quark level effective
Hamiltonian, and the Hamiltonian includes the $(S \pm iP)$ symmetry
mixing in operator
$O_{52}$, $O_{53}$, $O_{61}$ and $O_{62}$ in
eq.(\ref{eqn:hamil1}). Hence we consider the three
types of the internal interaction Hamiltonian for the
comparison. These Hamiltonians are given by
\begin{eqnarray}
  {\cal H}_{int}^{\Delta S=1} &=& {\cal H}^{min}_{int}\;\;
    \mbox{or}\;\; {\cal H}^{V-A}_{int}\;\;
      \mbox{or}\;\; {\cal H}^{S\pm iP}_{int}
  \label{eqn:effhamil03}\\
  {\cal H}_{int}^{min} &=& -\frac{G_f}{\sqrt{2}}\left(h_{D}
        \mbox{Tr}\bar{B}_v
       \left\{ h_{23},B_v\right\}
      +h_{F}\mbox{Tr}B_v
       \left[ h_{23},B_v\right]
     +h_{C}\bar{T}^{\mu}_v
    \left( h_{23}\right)
    T_{v\mu} \right)
   \label{eqn:operator02}
\end{eqnarray}
\begin{eqnarray}
  {\cal H}_{int}^{V-A} &=& -\frac{G_f}{\sqrt{2}}\left(h_{D}
        \mbox{Tr}\bar{B}_v
     \left\{\left(\xi^{\dag}h_{23}\xi\right),B_v\right\}
      +h_{F}\mbox{Tr}B_v
       \left[\left(\xi^{\dag}h_{23}\xi\right),B_v\right]
  \right.\nonumber\\
   &&\left.
    +h_{C}\bar{T}^{\mu}_v
    \left(\xi^{\dag}h_{23}\xi\right)
    T_{v\mu}
     +h_{M}\mbox{Tr}\left(\bar{B}_v B_v\right)\mbox{Tr}
        \left(\xi^{\dag}h_{23}\xi\right)
    +h_{N}\mbox{Tr}\left(\bar{T}^{\mu}_v T_{v\mu} \right)
        \left(\xi^{\dag}h_{23}\xi\right)\right)
   \nonumber\\
  \label{eqn:vaint}
\end{eqnarray}
\begin{eqnarray}
  {\cal H}_{int}^{S\pm iP} &=& -\frac{G_f}{\sqrt{2}}\left(h_{DS}
        \mbox{Tr}\bar{B}_v
       \left\{\left(\xi h_{23}\xi+\xi^{\dag}h_{23}\xi^
{\dag}\right),B_v\right\}
      +h_{FS}\mbox{Tr}B_v
       \left[\left(\xi h_{23}\xi+\xi^{\dag}h_{23}\xi^
{\dag}\right),B_v\right]
  \right.\nonumber\\
   &&+h_{MS}\mbox{Tr}\left(\bar{B}_v B_v\right)\mbox{Tr}
        \left(\xi h_{23}\xi+\xi^{\dag}h_{23}\xi^{\dag}\right)
    +h_{NS}\mbox{Tr}\left(\bar{T}^{\mu}_v T_{v\mu} \right)
        \left(\xi h_{23}\xi+\xi^{\dag}h_{23}\xi^{\dag}\right)
   \nonumber\\
   &&+h_{CS}\bar{T}^{\mu}_v
    \left(\xi h_{23}\xi+\xi^{\dag}h_{23}\xi^{\dag}\right)
    T_{v\mu}
  \label{eqn:spint}\\
         &&+h_{DP}
        \mbox{Tr}\bar{B}_v
       \left\{\left(\xi h_{23}\xi-\xi^{\dag}h_{23}\xi^
{\dag}\right),B_v\right\}
      +h_{FP}\mbox{Tr}B_v
       \left[\left(\xi h_{23}\xi-\xi^{\dag}h_{23}\xi^
{\dag}\right),B_v\right]
  \nonumber\\
   &&+h_{MP}\mbox{Tr}\left(\bar{B}_v B_v\right)\mbox{Tr}
        \left(\xi h_{23}\xi-\xi^{\dag}h_{23}\xi^{\dag}\right)
   +h_{NP}\mbox{Tr}\left(\bar{T}^{\mu}_v T_{v\mu} \right)
        \left(\xi h_{23}\xi-\xi^{\dag}h_{23}\xi^{\dag}\right)
   \nonumber\\
   && \left.
   +h_{CP}\bar{T}^{\mu}_v
    \left(\xi h_{23}\xi-\xi^{\dag}h_{23}\xi^{\dag}\right)
    T_{v\mu}\right).
   \nonumber
\end{eqnarray}
${\cal H}^{min}_{int}$ has the minimum elements for the internal
interaction and has three parameters $h_D$, $h_F$ and
$h_C$. ${\cal H}^{V-A}_{int}$ has the $V-A$ symmetry for the operator
$h_{23}$ and has five parameters.
${\cal H}^{S\pm iP}_{int}$ has
the symmetry mixing $(S\pm iP)$ for the operator $h_{23}$. The number
of parameters in this Hamiltonian is ten. In the following, using the
effective weak Hamiltonian (\ref{eqn:hamil1}) and the strong
interaction Lagrangian (\ref{eqn:strong1}) in Appendix, we
calculate hyperon non-leptonic weak decay amplitudes derived from
the tree diagrams and the one-loop diagrams. 
\setcounter{equation}{0}
\setcounter{figure}{0}
\section{Result and Discussion}
\subsection{Numerical analysis}

Ordinary non-leptonic weak decay amplitudes are represented by
\begin{equation}
  {\cal M}(B_i \longrightarrow B_f+\pi)
   =G_Fm_{{\pi}^+}^2\bar{u}_f(A+B\gamma_5)u_i.
  \label{eqn:ampform1}
\end{equation}
But the heavy baryon formalism of the chiral perturbation theory has
the following expression 
\begin{equation}
  {\cal M}(B_i \longrightarrow B_f+\pi)
   =\frac{G_f}{\sqrt{2}}
     \bar{U}_f({\cal A}+(q\cdot S_v){\cal B})U_i.
  \label{eqn:ampform2}
\end{equation}
Hence we use the transformation equations
\begin{equation}
  \begin{array}{lcl}
   A &=& \frac{-iG_f}{\sqrt{2}G_Fm_{{\pi}^+}^2}f_{\pi}{\cal A}
  \\
   B &=& \frac{iG_f}{2\sqrt{2}G_Fm_{{\pi}^+}^2}
         (E_f+m_f)f_{\pi}{\cal B},
  \end{array}
  \label{eqn:ampform3}
\end{equation}
and express the amplitudes in the ordinary form, which are also used
in ref.\cite{Takayama98}.
The amplitudes are calculated in the tree level and the chiral
logarithmic correction of the one-loop level with constants in
Table~3.1. 
The computational method is the same as in ref.\cite{Takayama98}.

The parameters in the effective Hamiltonian are fitted by the
experimental data and we can obtain hyperon weak decay
amplitudes. In the experimental data, the number of independent
data is eight. After the calculation of the diagrams, the
current-current interaction Hamiltonian
${\cal H}^{|\Delta S|=1}_{current}$  has unknown parameters which are
$a_1$, $a_2$, $a_3$ and $c_1$, and the numbers of unknown parameters
in the internal interaction Hamiltonians ${\cal H}_{int}^{min}$,
${\cal H}_{int}^{V-A}$ and ${\cal H}_{int}^{S\pm iP}$ are 3, 3 and 7,
respectively. Other parameters do not appear in the hyperon decay
amplitudes. The parameter $a_1$ and $a_2$ that appear in the scalar
current are determined as $a_1=29.1/m_s$ and $a_2=-94.8/m_s$, where
$m_s$ is the strange quark mass. They are derived from the baryon
mass difference with the chiral perturbation theory\cite{Jenkins92a}. 
Therefor the effective weak Hamiltonian
${\cal H}^{|\Delta S|=1}_{current}+{\cal H}^{min}_{int}$ and
${\cal H}^{|\Delta S|=1}_{current}+{\cal H}^{V-A}_{int}$ have
5 unknown parameters. It is possible to determine them by the
experimental data. But the effective Hamiltonian
${\cal H}^{|\Delta S|=1}_{current}+{\cal H}^{S\pm iP}_{int}$
has 9 unknown parameters and it is impossible to fix them
exactly. 
In the following we take two steps to fix unknown parameters.
First, introducing the assumption $h_{MP}=0.0$, the remaining 8
parameters are fitted by the experimental data. Next, the parameter
$h_{MP}$ is fitted with 8 experimental data. It is not sure that this
method gives the best fitted value for the decay amplitudes. In this
study, our aim is not to obtain the exact value of the decay
amplitudes, but is to obtain the parameter dependence of the amplitudes
and the characteristic of the effective weak Hamiltonian.

After the parameter
fitting, the obtained amplitudes and parameters are shown
in Table~3.1 and Table~3.2, respectively.
In these tables, the amplitudes ${\cal H}^{min}_{int}$ and
${\cal H}^{V-A(1)}_{int}$ are obtained by fitting 5 parameters in
the effective weak Hamiltonian 
${\cal H}^{|\Delta S|=1}_{current}+{\cal H}^{min}_{int}$
and
${\cal H}^{|\Delta S|=1}_{current}+{\cal H}^{V-A}_{int}$,
respectively. The amplitudes ${\cal H}^{V-A(2)}_{int}$ are obtained by 
fitting 7 parameters in the effective weak Hamiltonian
${\cal H}^{|\Delta S|=1}_{current}+{\cal H}^{V-A}_{int}$,
where the coupling constants $a_1$ and $a_2$ in
the scalar and pseudo-scalar currents are also considered as
unknown parameters. The amplitudes ${\cal H}^{S\pm iP}_{int}$ are
obtained by fitting 9 parameters, applying the above
assumptions. Table~3.2 suggests that the amplitudes
${\cal H}^{S\pm iP}_{int}$
and 
${\cal H}^{V-A(2)}_{int}$
reproduce the experimental data well. In
the following, we consider the amplitudes derived by 
${\cal H}^{S\pm iP}_{int}$
and
${\cal H}^{V-A(2)}_{int}$.
Table~3.4
and 3.5 show the components of the amplitudes derived by
${\cal H}^{S\pm iP}_{int}$
and
${\cal H}^{V-A(2)}_{int}$,
respectively. The tree level amplitudes caused by each operators in
eq. (\ref{eqn:hamil1}) are shown in Table~3.6 and 3.8. And the chiral
logarithmic corrections caused by each operator are shown in Table~3.7
and 3.9.

\subsection{Discussion}
It is known that the chiral perturbation theory cannot reproduce the
experimental data of hyperon non-leptonic weak decay
amplitudes\cite{Jenkins92b}. The effective weak Hamiltonians in the
previous study include the momentum independent baryon vertices, which 
correspond to the internal interaction Hamiltonian ${\cal H}^{V-A}_{int}$
in our analysis, and the momentum dependent meson weak vertices.
Considering the products of the quark currents for the weak
interaction, our effective Hamiltonian includes the momentum dependent 
baryon vertices and the effective Hamiltonian which depend on the
scalar and pseudo-scalar currents.
Our effective weak Hamiltonian is, therefore, the extension of the weak
Hamiltonian in ref.\cite{Jenkins92b}. Newly introduced terms have the
same form as in ref.\cite{Borasoy98}. But our method have the
constraints between the unknown parameters since we use the Noether
currents to construct the effective weak Hamiltonian.
In this paper we analyze the amplitudes caused by the internal
interaction ${\cal H}^{V-A}_{int}$ and ${\cal H}^{S\pm iP}_{int}$
since they reproduce the experimental data quite well.
The amplitudes derived by both Hamiltonians have the following
characteristics.
\begin{itemize}
  \begin{enumerate}
    \item   It is large that the contribution of the chiral logarithmic
	correction in the one-loop graphs, which is shown in Table~3.4 
	and 3.5.
    \item In the P-wave amplitudes, there are large cancelations
	between the amplitudes caused by the octet and the decuplet
        weak vertices in the logarithmic corrections.
    \item In the chiral logarithmic corrections, there are cancelations
	between the amplitudes caused by the operators $O_{61}$,
	$O_{62}$ and ${\cal H}^{|\Delta S|=1}_{int}$, which are shown
	in Table~3.7 and 3.9.
  \end{enumerate}
\end{itemize}
Our analysis suggests that the large cancelation improve the
theoretical prediction of the decay amplitudes. And it is important to 
consider the effective weak Hamiltonian caused by the scalar and the
pseudo-scalar currents.

Using the coupling constants $a_1$ and $a_2$ which derived by the
baryon mass difference in the chiral perturbation theory, it is
difficult to reproduce the experimental data with the internal
interaction Hamiltonian ${\cal H}^{V-A}_{int}$, which corresponds
to the amplitudes ${\cal H}^{V-A(1)}_{int}$ in Table~3.2.
If the constants $a_1$ and $a_2$ are considered as unknown
parameters and fitted by the experimentally obtained hyperon decay
amplitudes, the parameter $a_2$ becomes more than zero, which are
shown in Table~3.3, and it contradicts the calculation of the baryon
masses. In order to adopt the effective weak Hamiltonian
${\cal H}^{|\Delta S|=1}_{current}+{\cal H}^{V-A}_{int}$
for the hyperon weak decay, we have to consider the derivation of the
parameters $a_1$ and $a_2$.
\setcounter{equation}{0}
\setcounter{figure}{0}
\section{Conclusion}
Hyperon non-leptonic weak decays are analyzed with the chiral
perturbation theory. For the weak interaction which happens inside the 
hyperon and cannot be described by the products of the hadronic
currents, we introduce three types of the phenomenological weak
Hamiltonian. One includes only the two point vertex of the baryon
operators, and another includes $V-A$ type weak interaction and the
other includes $S\pm iP$ type weak interaction. Our analysis suggests
that the $V-A$ and the $S\pm iP$ type Hamiltonian can reproduce the
experimental data fairly well. We can get the outlook that hyperon 
non-leptonic weak decay amplitudes can be represented by the chiral
perturbation theory. But there remains the following three questions.
\begin{itemize}
  \begin{enumerate}
    \item $S\pm iP$ type Hamiltonian looks like exceed the framework
          of the standard theory. How does it realize the $S\pm iP$
          symmetry mixing in the weak interaction? 
    \item In the $V-A$ type Hamiltonian, the coupling constants $a_1$
          and $a_2$ have the same sign, which contradict the sign
          derived by the baryon mass. What cause such a contradiction?
          How can we get the exact values of $a_1$ and $a_2$?
    \item We use the constants in ref.\cite{Takayama98} for the
          numerical analysis. There is an ambiguity in the quark
          condensation value $B_0$. In order to reproduce the
          experimental data exactly, the $B_0$ value has to be
          decided. 
  \end{enumerate}
\end{itemize}

\appendix
\renewcommand{\theequation}{\arabic{equation}}
\setcounter{equation}{0}
{\bf Appendix.}\hspace{5mm}{\bf The Strong Interaction Lagrangian}
\vspace{5mm}

\hspace{5mm}
The strong interaction Lagrangian in the heavy baryon formalism in
the chiral perturbation theory is 
given by \cite{Jenkins91a,Jenkins91b}
\begin{eqnarray}
 {\cal L}_{strong} &=& \frac{f_{\pi}^2}{4}
    \mbox{Tr}\{({\cal D}_{\mu}\Sigma)({\cal D}^{\mu}\Sigma^{\dag})
    +\chi\Sigma^{\dag}+\Sigma\chi^{\dag} \}
  \nonumber  \\
   &&+ i\mbox{Tr}\bar{B}_v(v\cdot {\cal D})B_v
                   +2D\mbox{Tr}\bar{B}_v S^{\mu}_v\{A_{\mu},B_v\}
                   +2F\mbox{Tr}\bar{B}_v S^{\mu}_v[A_{\mu},B_v]
  \label{eqn:strong1}\\
  &&-i\bar{T}^{\mu}_v(v\cdot {\cal D})T_{v\mu}
      +{\cal C}\left(\bar{T}^{\mu}_v A_{\mu} B_v 
          +\bar{B}_v A_{\mu}T^{\mu}_v \right)
      +2{\cal H}\bar{T}^{\mu}_v S_{v\nu}A^{\nu}T_{v\mu}
  \nonumber\\
  &&+\Delta m \bar{T}^{\mu}_v T_{v\mu},
  \nonumber
\end{eqnarray}
where $A^{\mu}$, $V^{\mu}$ and ${\cal D}^{\mu}$ are represented by
\begin{eqnarray}
  A^{\mu}&=&
        \frac{i}{2}\left(\xi\partial^{\mu}\xi^{\dag}
         -\xi^{\dag}\partial^{\mu}\xi \right)
  \nonumber \\
  V^{\mu}&=&
        \frac{1}{2}\left(\xi\partial^{\mu}\xi^{\dag}
         +\xi^{\dag}\partial^{\mu}\xi \right)
  \nonumber \\
  {\cal D}^{\mu}B_v &=& \partial^{\mu}B_v+\left[V^{\mu},B_v\right].
  \nonumber
\end{eqnarray}
\newpage
{\bf\large Table~2.1} \hspace{3mm}
The values of the coefficients in the effective
weak Hamiltonian (\ref{eqn:weak1}). 
The values are taken from ref. \cite{Paschos90}.
 The data
set 1 corresponds to the choices $m_t=200$[GeV] and $\mu_0=0.24$[GeV],
$\Lambda_{QCD}=0.10$.
The data set 2 corresponds to the choices $m_t=200$[GeV] and
$\mu_0=0.71$[GeV], $\Lambda_{QCD}=0.316$. In both cases, $\mu_0$ is
defined so as to satisfy $\alpha_s(\mu^2)=1$.
\begin{center}
  \begin{tabular}{|c|c|c|}\hline
  &data set 1& data set 2 \\ \hline
  $\mu$(GeV)&0.24&0.71 \\ \hline
  $\Lambda^{(4)}$(GeV)&0.10&0.316 \\ \hline
  \hline
  $K_1$&-0.284&-0.270 \\ \hline
  $K_2$&0.009&0.011 \\ \hline
  $K_3$&0.026&0.027 \\ \hline
  $K_4$&0.026&0.027 \\ \hline
  $K_5$&0.004&0.002 \\ \hline
  $K_6$&0.004&0.002 \\ \hline
  \end{tabular}
\end{center}
\vspace{1cm}
\noindent{\bf\large Table~3.1} \hspace{3mm}
Hadron masses and coupling constants used in the numerical analysis of 
the hyperon decay amplitudes. $m$, $M$ and $f_{\pi}$ correspond  to
the baryon mass, the meson mass and pion decay constant,
respectively. Phenomenological coupling constants ${\cal D}$,
${\cal F}$ and ${\cal H}$ are taken from ref.
\cite{Jenkins91a,Jenkins91b}. The renormalized coupling constants
$L_4$ and $L_5$ are taken from ref.\cite{Gasser84}, where the
renormalization scale $4\pi\mu^2 =1.0\times 10^6\mbox{[MeV$^2$]}$ is
used. The constant $B_0$ corresponds to the quark condensation value
which is changed for the hyperon decay amplitudes\cite{Takayama98}. 
\[
  \begin{array}{ccccccccc}
  m_n &=& 939\mbox{[MeV]},&m_{\Sigma} &=& 1193\mbox{[MeV]},
  &m_{\Lambda} &=& 1116\mbox{[MeV]}\\
  m_{\Xi} &=& 1318\mbox{[MeV]},&m_{\Delta} &=& 1232\mbox{[MeV]},
  &m_{\Sigma^*} &=& 1385\mbox{[MeV]}\\
  m_{\Xi^*} &=& 1533\mbox{[MeV]}, &m_{\Omega} &=& 1672\mbox{[MeV]}
  &&&\\
  M_{\pi} &=& 138\mbox{[MeV]},& M_K &=& 496\mbox{[MeV]},
  &M_{\eta} &=& 547\mbox{[MeV]}\\
  D &=& 0.61,&F &=& 0.40,&{\cal C} &=& 1.6\\
  {\cal H} &=& -1.9, &f_{\pi} &=& 93\mbox{[Mev]}, 
  &L_4&=&-0.3003\\
  L_5&=&1.399,&B_0&=&139.14\mbox{[MeV]},&&&\\
  \end{array}
\]
\newpage
\noindent{\bf\large Table~3.2} \hspace{3mm}
The tree and one-loop level amplitudes.
Second column corresponds the experimentally obtained data.
Two types of the data set which is shown in Table~2.1 are used for the 
calculation of the S- and P-wave amplitudes.
{\small
\begin{center}
\begin{tabular}{|c|c|c|c|c|c|}
\multicolumn{6}{l}{{\normalsize S-wave:data set 1}}\\ \hline

process&exp.&${\cal H}_{int}^{min}$&${\cal H}_{int}^{V-A(1)}$
&${\cal H}_{int}^{V-A(2)}$&${\cal H}_{int}^{S\pm iP}$ \\ \hline
$\Sigma^-_-$  &$1.93\pm 0.01$  &5.803   &5.124   &2.486   &2.469
\\ \hline
$\Sigma^+_+$  &$0.06\pm 0.01$  &-0.1616 &-0.1537 &-0.1609 &-0.1838
\\ \hline 
$\Sigma^+_0$  &$-1.48\pm 0.05$ &-2.276  &-2.540  &-0.6831 &-0.7049
\\ \hline
$\Lambda^0_0$ &$-1.07\pm 0.02$ &0.9174  &0.1827  &-0.4942 &-0.4723
\\ \hline
$\Lambda^0_-$ &$1.47\pm 0.01$  &0.6009  &0.9013  &1.875   &1.883
\\ \hline
$\Xi^-_-$     &$-2.04\pm 0.01$ &-5.674  &-4.649  &-2.543  &-2.544
\\ \hline
$\Xi^0_0$     &$1.54\pm 0.03$  &2.089   &2.338   &0.8412  &0.8338
\\ \hline
\multicolumn{6}{l}{{\normalsize S-wave:data set 2}}\\ \hline
process&exp.&${\cal H}_{int}^{min}$&${\cal H}_{int}^{V-A(1)}$
&${\cal H}_{int}^{V-A(2)}$&${\cal H}_{int}^{S\pm iP}$ \\ \hline
$\Sigma^-_-$  &$1.93\pm 0.01$  &3.053    &3.455    &2.467    &2.460
\\ \hline
$\Sigma^+_+$  &$0.06\pm 0.01$  &-0.09894 &-0.09483 &-0.09799 &-0.1041
\\ \hline 
$\Sigma^+_0$  &$-1.48\pm 0.05$ &-1.124   &-1.405   &-0.7102  &-0.7155
\\ \hline
$\Lambda^0_0$ &$-1.07\pm 0.02$ &-0.2427  &-0.2096  &-0.4784  &-0.4669
\\ \hline
$\Lambda^0_-$ &$1.47\pm 0.01$  &1.555    &1.499    &1.887    &1.886
\\ \hline
$\Xi^-_-$     &$-2.04\pm 0.01$ &-3.152   &-3.393   &-2.535   &-2.534
\\ \hline
$\Xi^0_0$     &$1.54\pm 0.03$  &1.288    &1.463    &0.8527   &0.8478
\\ \hline
\end{tabular}
\end{center}
}
{\small
\begin{center}
\begin{tabular}{|c|c|c|c|c|c|}
\multicolumn{6}{l}{{\normalsize P-wave:data set 1}}\\ \hline

process&exp.&${\cal H}_{int}^{min}$&${\cal H}_{int}^{V-A(1)}$
&${\cal H}_{int}^{V-A(2)}$&${\cal H}_{int}^{S\pm iP}$ \\ \hline
$\Sigma^-_-$  &$-0.65\pm 0.07$ &1.400  &0.8779 &0.8624 &0.8623  
\\ \hline
$\Sigma^+_+$  &$19.07\pm 0.07$ &17.02  &17.50  &17.56  &17.56
\\ \hline 
$\Sigma^+_0$  &$12.04\pm 0.58$ &14.94  &14.12  &14.18  &14.18
\\ \hline
$\Lambda^0_0$ &$-7.14\pm 0.56$ &-9.935 &-7.240 &-7.237 &-7.237
\\ \hline
$\Lambda^0_-$ &$9.98\pm 0.24$  &8.004  &9.922  &9.912  &9.912
\\ \hline
$\Xi^-_-$     &$6.93\pm 0.31$  &8.796  &7.565  &7.569  &7.567
\\ \hline
$\Xi^0_0$     &$-6.43\pm 0.66$ &-3.792 &-5.507 &-5.526 &-5.529
\\ \hline
\multicolumn{6}{l}{{\normalsize P-wave:data set 2}}\\ \hline
process&exp.&${\cal H}_{int}^{min}$&${\cal H}_{int}^{V-A(1)}$
&${\cal H}_{int}^{V-A(2)}$&${\cal H}_{int}^{S\pm iP}$ \\ \hline
$\Sigma^-_-$  &$-0.65\pm 0.07$ &0.8957 &0.9010 &0.8955 &0.8957
\\ \hline
$\Sigma^+_+$  &$19.07\pm 0.07$ &17.52  &17.50  &17.52  &17.52
\\ \hline 
$\Sigma^+_0$  &$12.04\pm 0.58$ &14.23  &14.20  &14.23  &14.23
\\ \hline
$\Lambda^0_0$ &$-7.14\pm 0.56$ &-7.250 &-7.251 &-7.250 &-7.250
\\ \hline
$\Lambda^0_-$ &$9.98\pm 0.24$  &9.902  &9.906  &9.903  &9.902
\\ \hline
$\Xi^-_-$     &$6.93\pm 0.31$  &7.566  &7.568  &7.570  &7.566
\\ \hline
$\Xi^0_0$     &$-6.43\pm 0.66$ &-5.530 &-5.518 &-5.525 &-5.530
\\ \hline
\end{tabular}
\end{center}
}
\newpage
\vspace{8mm}
\noindent
{\bf\large Table~3.3} \hspace{3mm}
Parameters in the scalar and pseudo-scalar currents and in the
internal interaction Hamiltonian. 
Two types of data set are used for the parameter fitting.
The parameters are fitted with least square method for each effective
Hamiltonian.

\vspace{3mm}
\noindent
{\small
\begin{center}
\begin{tabular}{|c|c|c|}
\multicolumn{3}{l}{${\cal H}^{min}_{int}$}\\ \hline
parameter&data set 1&data set 2\\ \hline
\hline
$h_D$& -2.1603E-1&-1.5271E-1 \\ \hline
$h_F$&  4.4585E-1& 2.8016E-1 \\ \hline
$h_C$&  1.7336E 1& 1.1714E 1 \\ \hline
$a_1$&  1.9400E-1& 1.9400E-1 \\ \hline
$a_2$& -6.3200E-1&-6.3200E-1 \\ \hline
$a_3$&  4.0329E 1& 5.8615E 1 \\ \hline
$c_1$&  1.0971E 2& 1.0023E 2 \\ \hline
\end{tabular}
\begin{tabular}{|c|c|c|}
\multicolumn{3}{l}{${\cal H}^{V-A(1)}_{int}$}\\ \hline
parameter&data set 1&data set 2\\ \hline
\hline
$h_D$& -2.4929E-1&-1.9314E-1 \\ \hline
$h_F$&  5.1938E-1& 2.6855E-1 \\ \hline
$h_C$&  1.0896E 1& 8.4955    \\ \hline
$a_1$&  1.9400E-1& 1.9400E-1 \\ \hline
$a_2$& -6.3200E-1&-6.3200E-1 \\ \hline
$a_3$& -3.3276   &-3.6489    \\ \hline
$c_1$& -1.0240E 2&-2.5716E 2 \\ \hline
\end{tabular}
\end{center}
}
{\small
\begin{center}
\begin{tabular}{|c|c|c|}
\multicolumn{3}{l}{${\cal H}^{V-A(2)}_{int}$}\\ \hline
parameter&data set 1&data set 2\\ \hline
\hline
$h_D$& -8.6476E-2&-1.4807E-1 \\ \hline
$h_F$&  1.1019E-2& 4.5324E-2 \\ \hline
$h_C$&  1.0839E 1& 8.7308    \\ \hline
$a_1$&  6.2889E-2& 1.6273E-1 \\ \hline
$a_2$&  8.6041E-2& 3.6301E-2 \\ \hline
$a_3$& -9.4523   &-9.0925    \\ \hline
$c_1$& -2.4254E 2&-4.0113E 2 \\ \hline
\multicolumn{3}{l}{}\\
\multicolumn{3}{l}{}\\
\multicolumn{3}{l}{}\\
\multicolumn{3}{l}{}\\
\end{tabular}
\begin{tabular}{|c|c|c|}
\multicolumn{3}{l}{${\cal H}^{S \pm iP}_{int}$}\\ \hline
parameter&data set 1&data set 2\\ \hline
\hline
$h_{DS}$& -1.4405E-1   &-1.3792E-1    \\ \hline
$h_{FS}$&  2.9850E-1   & 2.0563E-1    \\ \hline
$h_{CS}$&  5.7787      & 4.7978       \\ \hline
$h_{DP}$& -1.2303E-1   &-6.4850E-3    \\ \hline
$h_{FP}$&  6.3642E-1   & 2.6492E-1    \\ \hline
$h_{CP}$& -1.4015      &-1.1624       \\ \hline
$h_{MP}$&  3.7366E-1   & 4.0328E-1    \\ \hline
$a_1$   &  1.9400E-1   & 1.9400E-1    \\ \hline
$a_2$   & -6.3200E-1   &-6.3200E-1    \\ \hline
$a_3$   & -1.5603      & 6.4024       \\ \hline
$c_1$   & -7.5060E 1   &-1.1388E 2    \\ \hline
\end{tabular}
\end{center}
}
\newpage
\vspace{8mm}
\noindent
{\bf\large Table~3.4}\hspace{3mm}
The predicted amplitudes with
the effective weak Hamiltonian 
${\cal H}^{|\Delta S|=1}_{current}+{\cal H}^{S\pm iP}_{int}$,
data set 1 and the parameters in
Table 3.2. $A_{exp}$  and $B_{exp}$ correspond to the experimental
data. $A_{tree}$ and $B_{tree}$ are the tree level amplitudes.
$\Delta A_{loop}$ and $\Delta B_{loop}$ are the summation of the chiral
logarithmic corrections of the one-loop graphs. They are represented
by $\Delta A_{loop}=\Delta A_{octet}+\Delta A_{decuplet}$ and
$\Delta B_{loop}=\Delta B_{octet}+\Delta B_{decuplet}$.
$A_{theory}$ and $B_{theory}$ are the total amplitude of the chiral
perturbation theory, which are represented by
$A_{theory}=A_{tree}+\Delta A_{loop}$ and
$B_{theory}=B_{tree}+\Delta B_{loop}$. 
{\footnotesize
\begin{center}
\begin{tabular}{|c|c|c|c|c|c|c|}
\multicolumn{7}{l}{{\normalsize S-wave:}}\\ \hline
decay mode&$A_{exp.}$&$A_{theory}$&$A_{tree}$&$\Delta A_{loop}$
&$\Delta A_{octet}$&$\Delta A_{decuplet}$\\ \hline
$\Sigma_-^-$ &$1.93\pm 0.01$
             &2.469   &-0.4684  &2.937  &2.905  &0.03211  \\ \hline
$\Sigma_+^+$ &$0.06\pm 0.01$ 
             &-0.1838&0.0     &-0.1838  &-0.06445&-0.1193 \\ \hline 
$\Sigma^+_0$ &$-1.48\pm 0.05$
             &-0.7049 &0.4723 &-1.177   &-1.964 &0.7863   \\ \hline
$\Lambda_0^0$&$-1.07\pm 0.02$
             &-0.4723 &0.4571 &-0.9293    &-0.7257 &-0.2036  \\ \hline
$\Lambda^0_-$&$1.47\pm 0.01$
             &1.883   &-0.4680  &2.351   &1.724  &0.6277  \\ \hline
$\Xi^-_-$    &$-2.04\pm 0.01$
             &-2.544  &0.5362 &-3.080    &-3.053  &-0.02747  \\ \hline
$\Xi^-_0$    &$1.54\pm 0.03$
             &0.8338   &-0.5217  &1.356   &1.909 &-0.5536   \\ \hline
\end{tabular}
\end{center}
}
{\footnotesize
\begin{center}
\begin{tabular}{|c|c|c|c|c|c|c|}
\multicolumn{7}{l}{{\normalsize P-wave:}}\\ \hline
decay mode&$B_{exp.}$&$B_{theory}$&$B_{tree}$&$\Delta B_{loop}$
&$\Delta B_{octet}$&$\Delta B_{decuplet}$\\ \hline
$\Sigma_-^-$ &$-0.65\pm 0.07$
             &0.8623 &0.9363  &-0.07397  &12.05 &-12.13  \\ \hline
$\Sigma_+^+$ &$19.07\pm 0.07$
             &17.56  &5.996  &11.56  &-12.37  &23.93  \\ \hline 
$\Sigma^+_0$ &$12.04\pm 0.58$
             &14.18  &3.857  &10.32  &-16.95 &27.27  \\ \hline
$\Lambda_0^0$&$-7.14\pm 0.56$
             &-7.237 &-2.674 &-4.563   &45.35  &-49.92  \\ \hline
$\Lambda^0_-$&$9.98\pm 0.24$
             &9.912  &2.401  &7.511  &-65.16 &72.67 \\ \hline
$\Xi^-_-$    &$6.93\pm 0.31$
             &7.567&-1.341   &8.909   &26.63  &-17.72  \\ \hline
$\Xi^-_0$    &$-6.43\pm 0.66$
             &-5.529 &1.327  &-6.856  &-18.20 &11.35 \\ \hline
\end{tabular}
\end{center}
}
\newpage
\vspace{8mm}
\noindent
{\bf\large Table~3.5}\hspace{3mm}
The predicted amplitudes with
the effective weak Hamiltonian
${\cal H}^{|\Delta S|=1}_{current}+{\cal H}^{V-A(2)}_{int}$
and data set 1.
$A_{exp}$  and $B_{exp}$ correspond to the experimental
data. $A_{tree}$ and $B_{tree}$ are the tree level amplitudes.
$\Delta A_{loop}$ and $\Delta B_{loop}$ are the summation of the chiral
logarithmic corrections of the one-loop graphs. They are represented
by $\Delta A_{loop}=\Delta A_{octet}+\Delta A_{decuplet}$ and
$\Delta B_{loop}=\Delta B_{octet}+\Delta B_{decuplet}$.
$A_{theory}$ and $B_{theory}$ are the total amplitude of the chiral
perturbation theory, which are represented by
$A_{theory}=A_{tree}+\Delta A_{loop}$ and
$B_{theory}=B_{tree}+\Delta B_{loop}$. 
{\footnotesize
\begin{center}
\begin{tabular}{|c|c|c|c|c|c|c|}
\multicolumn{7}{l}{{\normalsize S-wave:}}\\ \hline
decay mode&$A_{exp.}$&$A_{theory}$&$A_{tree}$&$\Delta A_{loop}$
&$\Delta A_{octet}$&$\Delta A_{decuplet}$\\ \hline
$\Sigma_-^-$ &$1.93\pm 0.01$
             &3.455    &0.05723  &3.398    &1.938    &1.460    \\ \hline
$\Sigma_+^+$ &$0.06\pm 0.01$ 
             &-0.09483 &0.0      &-0.09483 &-0.04009 &-0.05474 \\ \hline 
$\Sigma^+_0$ &$-1.48\pm 0.05$
             &-1.405   &0.1061   &-1.511   &-1.287   &-0.2243  \\ \hline
$\Lambda_0^0$&$-1.07\pm 0.02$
             &-0.2096  &0.03464  &-0.2442  &-0.8104  &0.5662   \\ \hline
$\Lambda^0_-$&$1.47\pm 0.01$
             &1.499    &0.1363   &1.363    &1.809    &-0.4460  \\ \hline
$\Xi^-_-$    &$-2.04\pm 0.01$
             &-3.393   &-0.08803 &-3.305   &-1.972   &-1.332   \\ \hline
$\Xi^-_0$    &$1.54\pm 0.03$
             &1.463    &-0.08577 &1.548    &1.178    &0.3707   \\ \hline
\end{tabular}
\end{center}
}
{\footnotesize
\begin{center}
\begin{tabular}{|c|c|c|c|c|c|c|}
\multicolumn{7}{l}{{\normalsize P-wave:}}\\ \hline
decay mode&$B_{exp.}$&$B_{theory}$&$B_{tree}$&$\Delta B_{loop}$
&$\Delta B_{octet}$&$\Delta B_{decuplet}$\\ \hline
$\Sigma_-^-$ &$-0.65\pm 0.07$
             &0.9010 &0.5866   &0.3144  &3.365  &-3.051  \\ \hline
$\Sigma_+^+$ &$19.07\pm 0.07$
             &17.50  &3.079    &14.42   &-5.204 &19.63   \\ \hline 
$\Sigma^+_0$ &$12.04\pm 0.58$
             &14.20  &2.053    &12.15   &-5.727 &17.88   \\ \hline
$\Lambda_0^0$&$-7.14\pm 0.56$
             &-7.251 &-1.001   &-6.250  &17.86  &-24.11  \\ \hline
$\Lambda^0_-$&$9.98\pm 0.24$
             &9.906  &-0.01803 &9.925   &-26.34 &36.26   \\ \hline
$\Xi^-_-$    &$6.93\pm 0.31$
             &7.568  &0.06998  &7.498   &11.24  &-3.737  \\ \hline
$\Xi^-_0$    &$-6.43\pm 0.66$
             &-5.518 &0.3436   &-5.862  &-7.270 &1.409   \\ \hline
\end{tabular}
\end{center}
}

\newpage
\vspace{8mm}
\noindent
{\bf\large Table~3.6}\hspace{3mm}
The amplitudes for each operator in the tree level
calculation with the effective weak Hamiltonian
${\cal H}^{|\Delta S|=1}_{current}+{\cal H}^{S\pm iP}_{int}$
and data set 1.
The first column corresponds to the operator types of the
effective Hamiltonian (\ref{eqn:weak1}) and
(\ref{eqn:spint}). The second row shows the
experimentally obtained data.\\
{\footnotesize
\begin{center}
\begin{tabular}{|c|c|c|c|c|c|c|c|}
\multicolumn{8}{l}{{\normalsize S-wave:}}\\ \hline
vertex&$\Sigma_-^-$&$\Sigma_+^+$&$\Sigma^+_0$&$\Lambda_0^0$&
$\Lambda^0_-$&$\Xi^-_-$&$\Xi^-_0$\\ \hline
exp.&1.93
&0.06
&-1.48
&-1.07
&1.47
&-2.04
&-1.54 \\ \hline 
$O_1$
& 0.2180
& 0.0
& -0.1542
& -0.1378
& 0.1949
& -0.2202
& 0.1557
\\ \hline
$O_2$
& 0.01382
& 0.0
& -0.009770
& -0.008735
& 0.01235
& -0.01396
& 0.009868
\\ \hline
$O_3$
& 0.01331
& 0.0
& -0.009408
& -0.008411
& 0.01190
& -0.01344
& 0.009502
\\ \hline
$O_4$
& 0.06653
& 0.0
& 0.09408
& 0.08411
& 0.05948
& -0.06719
& -0.09502
\\ \hline
$O_{51}$
& 0.0
& 0.0
& 0.0
& 0.0
& 0.0
& 0.0
& 0.0
\\ \hline
$O_{52}$
& 0.0007510
& 0.0
& -0.0005311
& -0.0004467
& 0.0006318
& -0.0007758
& 0.0005486
\\ \hline
$O_{53}$
& 0.0
& 0.0
& 0.0
& 0.0
& 0.0
& 0.0
& 0.0
\\ \hline
$O_{61}$
& -0.02103
& 0.0
& 0.01487
& 0.01251
& -0.01769
& 0.02172
& -0.01536
\\ \hline
$O_{62}$
& 0.0
& 0.0
& 0.0
& 0.0
& 0.0
& 0.0
& 0.0
\\ \hline
${\cal H}^{S\pm iP}_{int}$
& -0.7598
& 0.0
& 0.5372
& 0.5159
& -0.7295
& 0.8300
& -0.5869
\\ \hline
\end{tabular}
\end{center}
}
{\footnotesize
\begin{center}
\begin{tabular}{|c|c|c|c|c|c|c|c|}
\multicolumn{8}{l}{{\normalsize P-wave:}}\\ \hline
vertex&$\Sigma_-^-$&$\Sigma_+^+$&$\Sigma^+_0$&$\Lambda_0^0$&
$\Lambda^0_-$&$\Xi^-_-$&$\Xi^-_0$\\ \hline
exp.&-0.65
&19.07
&12.04
&-7.14
&9.98
&6.93
&-6.43 \\ \hline 
$O_1$
& 0.4317
& 0.0
& -0.3053
& 1.066
& -1.508
& 0.5847
& -0.4135
\\ \hline
$O_2$
& 0.02736
& 0.0
& -0.01935
& 0.06759
& -0.09559
& 0.03706
& -0.02621
\\ \hline
$O_3$
& 0.02635
& 0.0
& -0.01863
& 0.06509
& -0.09205
& 0.03569
& -0.02523
\\ \hline
$O_4$
& 0.1317
& 0.0
& 0.1863
& -0.6509
& -0.4602
& 0.1784
& 0.2523
\\ \hline
$O_{51}$
& 0.0
& 0.0
& 0.0
& 0.0
& 0.0
& 0.0
& 0.0
\\ \hline
$O_{52}$
& 0.001240
& 0.01128
& 0.007101
& -0.006488
& 0.009176
& -0.005211
& 0.003685
\\ \hline
$O_{53}$
& -0.001368
& 0.0
& 0.0009674
& -0.003380
& 0.004779
& -0.001853
& 0.001310
\\ \hline
$O_{61}$
& -0.03472
& -0.3159
& -0.1988
& 0.1817
& -0.2569
& 0.1459
& -0.1032
\\ \hline
$O_{62}$
& 0.03831
& 0.0
& -0.02709
& 0.09463
& -0.1338
& 0.05188
& -0.03669
\\ \hline
${\cal H}^{S\pm iP}_{int}$
& 0.3157
& 6.301
& 4.232
& -3.489
& 4.934
& -2.368
& 1.675
\\ \hline
\end{tabular}
\end{center}
}

\newpage
\vspace{8mm}
\noindent
{\bf\large Table~3.7}\hspace{3mm}
The amplitudes for each operator, which are derived from the one-loop
calculation with the effective weak Hamiltonian
${\cal H}^{|\Delta S|=1}_{current}+{\cal H}_{int}^{S\pm iP}$
and data set 1.
The first column corresponds to the operator types of the
effective Hamiltonian (\ref{eqn:weak1}) and
(\ref{eqn:spint}). The second row shows the
experimentally obtained data.\\
{\footnotesize
\begin{center}
\begin{tabular}{|c|c|c|c|c|c|c|c|}
\multicolumn{8}{l}{{\normalsize S-wave:}}\\ \hline
vertex&$\Sigma_-^-$&$\Sigma_+^+$&$\Sigma^+_0$&$\Lambda_0^0$&
$\Lambda^0_-$&$\Xi^-_-$&$\Xi^-_0$\\ \hline
exp.&1.93
&0.06
&-1.48
&-1.07
&1.47
&-2.04
&-1.54 \\ \hline 
$O_1$
& 0.8530
& 0.03913
& -0.5675
& -0.8295
& 1.182
& -1.275
& 0.9032
\\ \hline
$O_2$
& 0.05406
& 0.002480
& -0.03597
& -0.05257
& 0.07492
& -0.08080
& 0.05724
\\ \hline
$O_3$
& 0.1296
& 0.004146
& -0.08691
& -0.05800
& 0.08250
& -0.1622
& 0.1134
\\ \hline
$O_4$
& 0.3841
& -0.07977
& 0.4795
& 0.4600
& 0.3195
& -0.3399
& -0.4790
\\ \hline
$O_{51}$
& 0.0
& 0.0
& 0.0
& 0.0
& 0.0
& 0.0
& 0.0
\\ \hline
$O_{52}$
& -0.01267
& 0.00001678
& 0.008974
& 0.05773
& -0.08168
& 0.02882
& -0.02036
\\ \hline
$O_{53}$
& -0.1266
& 0.004708
& 0.08431
& 0.06087
& -0.08676
& 0.1230
& -0.08344
\\ \hline
$O_{61}$
& 0.3548
& -0.0004700
& -0.2513
& -1.616
& 2.287
& -0.8070
& 0.5701
\\ \hline
$O_{62}$
& 3.545
& -0.1318
& -2.361
& -1.704
& 2.429
& -3.444
& 2.336
\\ \hline
${\cal H}^{S\pm iP}_{int}$
& -2.244
& -0.02220
& 1.552
& 2.753
& -3.856
& 2.877
& -2.042
\\ \hline
\end{tabular}
\end{center}
}
{\footnotesize
\begin{center}
\begin{tabular}{|c|c|c|c|c|c|c|c|}
\multicolumn{8}{l}{{\normalsize P-wave:}}\\ \hline
vertex&$\Sigma_-^-$&$\Sigma_+^+$&$\Sigma^+_0$&$\Lambda_0^0$&
$\Lambda^0_-$&$\Xi^-_-$&$\Xi^-_0$\\ \hline
exp.&-0.65
&19.07
&12.04
&-7.14
&9.98
&6.93
&-6.43 \\ \hline 
$O_1$
& 0.5927
& 7.247
& 4.705
& -1.630
& 2.305
& 0.008264
& -0.005844
\\ \hline
$O_2$
& 0.04233
& 0.4647
& 0.2987
& -0.1014
& 0.1434
& 0.004622
& -0.003268
\\ \hline
$O_3$
& 0.06285
& -0.2282
& -0.2058
& 0.1922
& -0.2718
& -0.1290
& 0.09123
\\ \hline
$O_4$
& 0.9896
& -0.004056
& 1.397
& 0.4862
& 0.3438
& -0.2587
& -0.3658
\\ \hline
$O_{51}$
& -7.953E-4
& -8.974E-4
& -7.224E-5
& -3.121E-4
& 4.414E-4
& -6.831E-4
& 4.830E-4
\\ \hline
$O_{52}$
& -0.8894
& -0.3386
& 0.3895
& -1.813
& 2.564
& -1.055
& 0.7461
\\ \hline
$O_{53}$
& 0.1745
& 0.7492
& 0.4066
& -0.3720
& 0.5251
& -0.3250
& 0.2301
\\ \hline
$O_{61}$
& 24.90
& 9.480
& -10.91
& 50.76
& -71.79
& 29.55
& -20.89
\\ \hline
$O_{62}$
& -4.885
& -20.98
& -11.39
& 10.42
& -14.70
& 9.100
& -6.443
\\ \hline
${\cal H}^{S\pm iP}_{int}$
& -21.06
& 15.17
& 25.62
& -62.50
& 88.39
& -27.98
& 19.79
\\ \hline
\end{tabular}
\end{center}
}

\newpage
\vspace{8mm}
\noindent
{\bf\large Table~3.8}\hspace{3mm}
The amplitudes for each operator in the tree level
calculation with the effective weak Hamiltonian
${\cal H}^{|\Delta S|=1}_{current}+{\cal H}^{V-A}_{int}$
and data set 1.
The first column corresponds to the operator types of the
effective Hamiltonian (\ref{eqn:weak1}) and
(\ref{eqn:vaint}). The second row shows the
experimentally obtained data.\\
{\footnotesize
\begin{center}
\begin{tabular}{|c|c|c|c|c|c|c|c|}
\multicolumn{8}{l}{{\normalsize S-wave:}}\\ \hline
vertex&$\Sigma_-^-$&$\Sigma_+^+$&$\Sigma^+_0$&$\Lambda_0^0$&
$\Lambda^0_-$&$\Xi^-_-$&$\Xi^-_0$\\ \hline
exp.&1.93
&0.06
&-1.48
&-1.07
&1.47
&-2.04
&-1.54 \\ \hline 
$O_1$
& 0.2180
& 0.0
& -0.1542
& -0.1378
& 0.1949
& -0.2202
& 0.1557
\\ \hline
$O_2$
& 0.01382
& 0.0
& -0.009770
& -0.008735
& 0.01235
& -0.01396
& 0.009868
\\ \hline
$O_3$
& 0.01331
& 0.0
& -0.009408
& -0.008411
& 0.01190
& -0.01344
& 0.009502
\\ \hline
$O_4$
& 0.06653
& 0.0
& 0.09408
& 0.08411
& 0.05948
& -0.06719
& -0.09502
\\ \hline
$O_{51}$
& 0.0
& 0.0
& 0.0
& 0.0
& 0.0
& 0.0
& 0.0
\\ \hline
$O_{52}$
& -2.105E-5
& 0.0
& 1.489E-5
& 8.426E-5
& -1.192E-4
& 7.247E-5
& -5.124E-5
\\ \hline
$O_{53}$
& 0.0
& 0.0
& 0.0
& 0.0
& 0.0
& 0.0
& 0.0
\\ \hline
$O_{61}$
& 0.0005894
& 0.0
& -0.0004168
& -0.002359
& 0.003336
& -0.002029
& 0.001435
\\ \hline
$O_{62}$
& 0.0
& 0.0
& 0.0
& 0.0
& 0.0
& 0.0
& 0.0
\\ \hline
${\cal H}^{V-A}_{int}$
& -0.05072
& 0.0
& 0.03587
& -0.008023
& 0.01135
& 0.02539
& -0.01795
\\ \hline
\end{tabular}
\end{center}
}
{\footnotesize
\begin{center}
\begin{tabular}{|c|c|c|c|c|c|c|c|}
\multicolumn{8}{l}{{\normalsize P-wave:}}\\ \hline
vertex&$\Sigma_-^-$&$\Sigma_+^+$&$\Sigma^+_0$&$\Lambda_0^0$&
$\Lambda^0_-$&$\Xi^-_-$&$\Xi^-_0$\\ \hline
exp.&-0.65
&19.07
&12.04
&-7.14
&9.98
&6.93
&-6.43 \\ \hline 
$O_1$
& 0.4317
& 0.0
& -0.3053
& 1.066
& -1.508
& 0.5847
& -0.4135
\\ \hline
$O_2$
& 0.02736
& 0.0
& -0.01935
& 0.06759
& -0.09559
& 0.03706
& -0.02621
\\ \hline
$O_3$
& 0.02635
& 0.0
& -0.01863
& 0.06509
& -0.09205
& 0.03569
& -0.02523
\\ \hline
$O_4$
& 0.1317
& 0.0
& 0.1863
& -0.6509
& -0.4602
& 0.1784
& 0.2523
\\ \hline
$O_{51}$
& 0.0
& 0.0
& 0.0
& 0.0
& 0.0
& 0.0
& 0.0
\\ \hline
$O_{52}$
& -4.622E-4
& -8.578E-4
& -2.797E-4
& 1.235E-3
& -1.746E-3
& 1.528E-3
& -1.080E-3
\\ \hline
$O_{53}$
& -0.001368
& 0.0
& 0.0009674
& -0.003380
& 0.004779
& -0.001853
& 0.001310
\\ \hline
$O_{61}$
& 0.01294
& 0.02402
& 0.007832
& -0.03457
& 0.04890
& -0.04278
& 0.03025
\\ \hline
$O_{62}$
& 0.03831
& 0.0
& -0.02709
& 0.09463
& -0.1338
& 0.05188
& -0.03669
\\ \hline
${\cal H}^{V-A}_{int}$
& -0.2121
& 0.4736
& 0.4848
& -0.04624
& 0.06539
& 0.2913
& -0.2060
\\ \hline
\end{tabular}
\end{center}
}

\newpage
\vspace{8mm}
\noindent
{\bf\large Table~3.9}\hspace{3mm}
The amplitudes for each operator, which are derived from the one-loop
calculation with the effective weak Hamiltonian
${\cal H}^{|\Delta S|=1}_{current}+{\cal H}^{V-A}_{int}$
and data set 1.
The first column corresponds to the operator types of the
effective Hamiltonian (\ref{eqn:weak1}) and
(\ref{eqn:vaint}). The second row shows the
experimentally obtained data.\\
{\footnotesize
\begin{center}
\begin{tabular}{|c|c|c|c|c|c|c|c|}
\multicolumn{8}{l}{{\normalsize S-wave:}}\\ \hline
vertex&$\Sigma_-^-$&$\Sigma_+^+$&$\Sigma^+_0$&$\Lambda_0^0$&
$\Lambda^0_-$&$\Xi^-_-$&$\Xi^-_0$\\ \hline
exp.&1.93
&0.06
&-1.48
&-1.07
&1.47
&-2.04
&-1.54 \\ \hline 
$O_1$
& 0.8530
& 0.03913
& -0.5675
& -0.8295
& 1.182
& -1.275
& 0.9032
\\ \hline
$O_2$
& 0.05406
& 0.002480
& -0.03597
& -0.05257
& 0.07492
& -0.08080
& 0.05724
\\ \hline
$O_3$
& 0.1296
& 0.004146
& -0.08691
& -0.05800
& 0.08250
& -0.1622
& 0.1134
\\ \hline
$O_4$
& 0.3841
& -0.07977
& 0.4795
& 0.4600
& 0.3195
& -0.3399
& -0.4790
\\ \hline
$O_{51}$
& 0.0
& 0.0
& 0.0
& 0.0
& 0.0
& 0.0
& 0.0
\\ \hline
$O_{52}$
& -0.04769
& -2.247
& 0.03372
& 0.1913
& -0.2706
& 0.1004
& -0.07101
\\ \hline
$O_{53}$
& -0.02115
& 0.004708
& 0.009744
& -0.01165
& 0.01580
& 0.007148
& -0.001518
\\ \hline
$O_{61}$
& 1.335
& 0.00006291
& -0.9442
& -5.357
& 7.576
& -2.812
& 1.988
\\ \hline
$O_{62}$
& 0.5923
& -0.1318
& -0.2728
& 0.3262
& -0.4424
& -0.2001
& 0.04251
\\ \hline
${\cal H}^{V-A}_{int}$
& -1.055
& 0.0001877
& 0.7451
& 4.919
& -6.956
& 2.511
& -1.776
\\ \hline
\end{tabular}
\end{center}
}
{\footnotesize
\begin{center}
\begin{tabular}{|c|c|c|c|c|c|c|c|}
\multicolumn{8}{l}{{\normalsize P-wave:}}\\ \hline
vertex&$\Sigma_-^-$&$\Sigma_+^+$&$\Sigma^+_0$&$\Lambda_0^0$&
$\Lambda^0_-$&$\Xi^-_-$&$\Xi^-_0$\\ \hline
exp.&-0.65
&19.07
&12.04
&-7.14
&9.98
&6.93
&-6.43 \\ \hline 
$O_1$
& 0.5927
& 7.247
& 4.705
& -1.630
& 2.305
& 0.008264
& -0.005844
\\ \hline
$O_2$
& 0.04233
& 0.4647
& 0.2987
& -0.1014
& 0.1434
& 0.004622
& -0.003268
\\ \hline
$O_3$
& 0.06285
& -0.2282
& -0.2058
& 0.1922
& -0.2718
& -0.1290
& 0.09123
\\ \hline
$O_4$
& 0.9896
& -0.004056
& 1.397
& 0.4862
& 0.3438
& -0.2587
& -0.3658
\\ \hline
$O_{51}$
& -7.953E-4
& -8.974E-4
& -7.224E-5
& -3.121E-4
& 4.414E-4
& -6.831E-4
& 4.830E-4
\\ \hline
$O_{52}$
& -1.617
& -1.384
& 0.1645
& -2.453
& 3.469
& -1.508
& 1.067
\\ \hline
$O_{53}$
& 0.6537
& -0.1189
& -0.5461
& 1.436
& -2.032
& 0.9439
& -0.6671
\\ \hline
$O_{61}$
& 45.27
& 38.76
& -4.605
& 68.69
& -97.14
& 42.24
& -29.87
\\ \hline
$O_{62}$
& -18.30
& 3.329
& 15.29
& -40.22
& 56.90
& -26.43
& 18.68
\\ \hline
${\cal H}^{V-A}_{int}$
& -27.28
& -31.00
& -2.629
& -34.20
& 48.36
& -8.436
& 5.969
\\ \hline
\end{tabular}
\end{center}
}
\newpage
\setcounter{section}{2}
\setcounter{figure}{0}
\begin{figure}
    \centerline{\epsfxsize=130mm \epsfbox{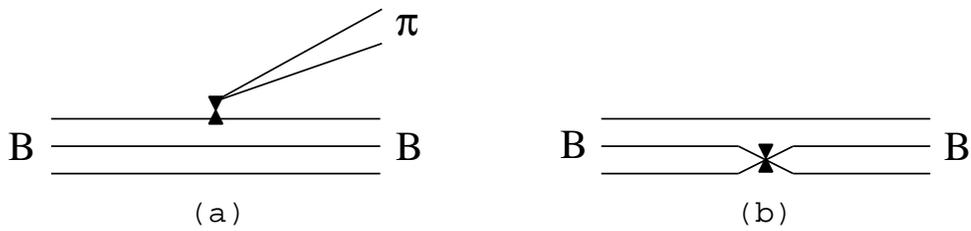}}
    \caption{$\Delta S=1$ hyperon non-leptonic weak decay diagrams of
             quark currents.
             (a) corresponds to the weak interaction between two
             hadronic currents which are color singlet.
             (b) corresponds to the weak interaction inside the hyperon.} 
    \label{fig:quarks}
\end{figure}
\end{document}